\documentclass[journal,twoside,web]{ieeecolor}
\usepackage{amsmath,amsfonts}
\usepackage{algorithmic}
\usepackage{array}
\usepackage[caption=false,font=normalsize,labelfont=sf,textfont=sf]{subfig}
\usepackage{textcomp}
\usepackage{stfloats}
\usepackage{url}
\usepackage{verbatim}
\usepackage{graphicx}
\usepackage{booktabs}
\usepackage{multirow}
\usepackage{tabularx}
\usepackage{bm}
\usepackage{cite}
\usepackage[T1]{fontenc}
\usepackage{balance}
\usepackage{generic}
\usepackage{cite}
\usepackage{amsmath,amssymb,amsfonts}
\usepackage{algorithmic}
\usepackage{graphicx}
\usepackage{algorithm,algorithmic}
\usepackage{hyperref}
\hypersetup{hidelinks=true}
\usepackage{textcomp}
\def\BibTeX{{\rm B\kern-.05em{\sc i\kern-.025em b}\kern-.08em
    T\kern-.1667em\lower.7ex\hbox{E}\kern-.125emX}}
\begin{document}
\title{SP-DiffDose: A Conditional Diffusion Model for Radiation Dose Prediction Based on Multi-Scale Fusion of Anatomical Structures, Guided by SwinTransformer and Projector}
\author{Linjie Fu,  Xia Li, Xiuding Cai, Yingkai Wang, Xueyao Wang, Yu Yao, and Yali Shen*
\thanks{Linjie Fu, Xiuding Cai, Yingkai Wang, Xueyao Wang, Yu Yao are with Chengdu Computer Application Institute Chinese Academy of Sciences, China and University of the Chinese Academy of Sciences, China.}
\thanks{Xia Li are with Radiophysical Technology Center, Cancer Center, West China Hospital, Sichuan University, China.}
\thanks{Yali Shen are with Sichuan University West China Hospital Department of Abdominal Oncology, China.}
\thanks{* is the corresponding author. Email:sylprecious123@163.com.}
\thanks{This work was supported by 1.3.5 project for disciplines of excellence, West China Hospital, Sichuan University(20HXJS040).}}

\maketitle
\begin{abstract}
Radiation therapy serves as an effective and standard method for cancer treatment. Excellent radiation therapy plans always rely on high-quality dose distribution maps obtained through repeated trial and error by experienced experts. However, due to individual differences and complex clinical situations, even seasoned expert teams may need help to achieve the best treatment plan every time quickly. Many automatic dose distribution prediction methods have been proposed recently to accelerate the radiation therapy planning process and have achieved good results. However, these results suffer from over-smoothing issues, with the obtained dose distribution maps needing more high-frequency details, limiting their clinical application. To address these limitations, we propose a dose prediction diffusion model based on SwinTransformer and a projector, SP-DiffDose. To capture the direct correlation between anatomical structure and dose distribution maps, SP-DiffDose uses a structural encoder to extract features from anatomical images, then employs a conditional diffusion process to blend noise and anatomical images at multiple scales and gradually map them to dose distribution maps. To enhance the dose prediction distribution for organs at risk, SP-DiffDose utilizes SwinTransformer in the deeper layers of the network to capture features at different scales in the image. To learn good representations from the fused features, SP-DiffDose passes the fused features through a designed projector, improving dose prediction accuracy. Finally, we evaluate SP-DiffDose on an internal dataset. The results show that SP-DiffDose outperforms existing methods on multiple evaluation metrics, demonstrating the superiority and generalizability of our method.
\end{abstract}


\section{Introduction}
\begin{figure}
\centering
\includegraphics[width=1\linewidth]{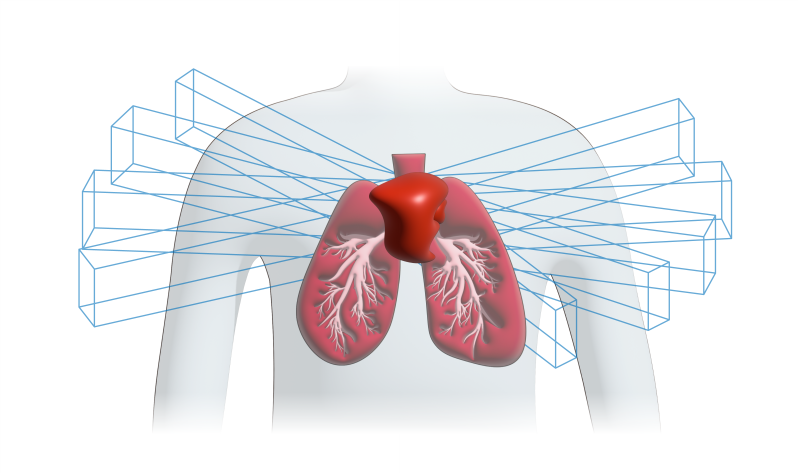}
   \caption{Demonstrate a radiation therapy plan using beam-shaped radiation.}
\label{fig1}
\end{figure}

\IEEEPARstart{R}{adiation} therapy, as one of the primary modalities for cancer treatment, relies on accurate and personalized treatment plans to control tumors maximally while protecting surrounding healthy tissues\cite{men2007exact}. With the support of modern radiation therapy techniques such as Intensity-Modulated Radiation Therapy (IMRT), cancer treatment outcomes have significantly improved over the past decades\cite{lee2018intensity}. However, formulating a high-quality radiation therapy plan remains a challenge, especially when dealing with tumors characterized by complex geometric shapes and positions, such as lung cancer. In these situations, a conflict exists between the planning target volume (PTV) and the organs at risk (OARs), often requiring radiation to be delivered from multiple different directions to achieve treatment (Fig.~\ref{fig1}). Moreover, developing a radiation therapy plan in clinical practice requires significant collaboration between medical physicists and radiation oncologists. They communicate and adjust the plan multiple times to balance therapeutic goals and minimize harm to healthy tissues. Medical physicists design and optimize the plan, while radiation oncologists assess its clinical suitability\cite{braam2006intensity}. The plan often undergoes several modifications, which can be time-consuming and potentially delay treatment. Even experienced teams can struggle to quickly devise the best treatment plan due to individual differences and complex clinical situations. Therefore, it is essential to develop a new method for automated dose prediction to quickly generate optimized treatment plans, improving the efficiency and consistency of treatment planning while reducing the workload of clinical professionals, thereby enhancing overall treatment efficiency and ensuring treatment quality.

Knowledge-Based Planning (KBP) is a traditional dose prediction method. KBP primarily relies on the data from historical treatment plans, predicting an optimal treatment plan for new patients by analyzing patterns and regularities within this data\cite{wu2011data}. However, KBP depends on the quality and quantity of the available treatment plan data. If these data are representative and sufficient in quantity, the accuracy and reliability of the prediction can be maintained. In recent years, due to the superior capability of deep learning in handling complex and non-standardized data, many Deep Neural Networks (DNNs)  have been proposed for dose prediction\cite{nguyen2017dose,tan2021incorporating,liu2021cascade,wang2022vmat,song2020dose,mahmood2018automated,jhanwar2022domain,jiao2023transdose}. For instance, Liu et al.\cite{liu2021cascade} construct a Cascaded 3D (C3D) UNet that integrates global and local anatomic features for fine-tuned dose prediction. Jiao et al.\cite{jiao2023transdose} use Graph Convolutional Networks (GCN) to extract specific category features from CT images, replacing the input of anatomical structures for dose prediction.

Although DNNs have achieved satisfactory results in dose distribution prediction, they share a common flaw. DCNNs always apply L1 or L2 loss fuction to guide model optimization. These loss functions compute the posterior mean of the joint distribution between the prediction and the actual data, resulting in excessively smoothed outcomes that miss crucial high-frequency details\cite{xie2023diffusion}. As illustrated in Fig.~\ref{fig3}, compared to ground truth, the images predicted by DNNs (Columns three and four) lack details in the direction of the radiation beam and information about the beam intensity. This information is crucial for clinical assessment of the quality of treatment plans. Therefore, there is a need to explore a method that allows the network to predict high-frequency details.
In recent years, diffusion model have excelled in many tasks, proving their colossal potential in modeling complex image distributions\cite{ho2020denoising,nichol2021improved,choi2021ilvr}. In contrast to DCNNs, the diffusion model solely requires the prediction of the noise introduced during the forward process in training, thereby eliminating the need for any extra assumptions about the target data distribution and preventing over-smoothing. Feng et al.\cite{feng2023diffdp} propose using the diffusion model for dose prediction, and the results show that it is closer to the ground truth regarding the direction and intensity of the radiation beam.

Considering the excellent performance of the diffusion model, we propose SP-DiffDose, an end-to-end dose prediction method based on the diffusion model. SP-DiffDose has two network branches. The first is the denoising network, used to restore pure Gaussian noise into a dose distribution map. The second is the structural encoding network, which extracts anatomical structure features from CT images, PTV, and OARs to auxiliary the denoising network in restoring the dose distribution map. Specifically, the denoising network combines convolution and transformer; we use convolution layers at the beginning to gradually extract and upgrade features, then use the SwinTransformer blocks\cite{liu2021swin} to capture local and global features. This progressive feature refinement aids in establishing the abstraction of information from simple to complex. The structural encoding network is similar to the denoising network but only includes the encoder part. Lastly, to further infuse the anatomical information into the denoising network, we design a projector to optimize the fused features.

We can summarize the novelty and contributions of our work in four aspects: 1) We propose an end-to-end network SP-DiffDose based on the diffusion model to predict dose distribution, addressing the issue of excessive smoothing in existing methods. 2) We design a network architecture combining convolution and SwinTransformer to capture local and global features for more accurate predictions. 3) We develop a structural encoder to extract available anatomical information from CT images, PTVs, and OARs and introduce a projector to guide the denoising network for more precise predictions. 4) Experimental results from the chest tumor dataset demonstrate the superiority and generalizability of our method.

\section{Related work}
\subsection{Dose distribution prediction}
To automatically predict the dose distribution, many methods utilize high-quality radiation therapy plan databases to establish predictive models. Standard methods include KBP and DCNNs algorithms.

Knowledge-Based Planning (KBP) algorithms measure similarity to find treatment plans from a database that matches the current case\cite{shiraishi2016knowledge,ge2019knowledge}. These plans then inform the treatment plan for a new patient. For example, schreibmann et al.\cite{schreibmann2015automated} use an iterative nearest neighbor point registration algorithm to find similar treatment plans, transferring parameters like field settings and multi-leaf collimator positions to improve the treatment plan. However, KBP have limitations. They require manual extraction of anatomical structure features, which may not be comprehensive enough for complex cancer types, leading to less accurate predictions. Additionally, KBP must fit relationships between anatomical structures and dose distribution, which is often less effective for complex nonlinear relationships.

With the rapid development of Deep Learning (DL)\cite{lecun2015deep}, it is capability to automatically extract image features has led to its widespread application in fields like computer vision and medical image analysis. Therefore, using DL for dose prediction has become a current research focus. For example, Kearney et al.\cite{kearney2018dosenet} design DoseNet for predicting 3D dose distribution in prostate radiation therapy plans. To further improve prediction accuracy and computational efficiency, Nguyen et al.\cite{nguyen2017dose} introduce HD-UNet for dose distribution prediction. HD-UNet is based on 3D U-Net and DenseNet, using dense connections to concatenate feature maps of different levels and network depths, reducing the loss of features during downsampling steps and enhancing prediction accuracy. Furthermore, Jhanwar et al.\cite{jhanwar2022domain} introduce a moment-based loss function, incorporating Dose-Volume Histograms (DVH) as domain knowledge into 3D dose prediction without computational overhead and translating it into deliverable Pareto optimal plans. Wang et al.\cite{wang2022deep} propose a novel "decompose-assemble" strategy and a global-to-beam framework, learning the entire dose image space, decomposing it into beam-based subfractions, and using a multi-beam voting strategy for the final dose map. Jiao et al.\cite{jiao2023transdose} introduce a method based on Transformer and Graph Convolutional Network (GCN) for predicting radiation therapy dose distribution from CT images. It includes a superpixel-level GCN classification branch for extracting anatomical information, combined with Transformer modules and U-net architecture, achieving accurate dose distribution prediction. Although these methods have shown good performance on objective metrics, they all exhibit over-smoothing in visualization results. It is because these method use L1 or L2 loss to optimize predictions and ground truth, which is very sensitive to outliers and tries to predict all samples to minimize squared error, resulting in a smoothed effect. To address this issue, Feng et al.\cite{feng2023diffdp} propose DiffDP, progressively adding noise and training a noise predictor to convert the dose distribution map into Gaussian noise gradually, then progressively removing the noise using the trained noise predictor, thereby solving the over-smoothing problem and improving the accuracy and precision of radiation therapy dose prediction. Inspired by DiffDP, we propose SP-DiffDose, achieving better performance than state-of-the-art methods.
\subsection{Diffusion model}
Recently, diffusion model have shown excellent performance in creating high-quality images\cite{dhariwal2021diffusion,rombach2022high,yang2022diffusion}. Researchers are now applying diffusion model to various medical imaging tasks, including classification\cite{yang2023diffmic,han2022card}, segmentation\cite{amit2021segdiff,wolleb2022diffusion}, denoising\cite{zhao2023diffuld}, and synthesis\cite{li2023zero,lyu2022conversion,ozbey2023unsupervised}. Ozbey et al.\cite{ozbey2023unsupervised} introduce SynDiff, which achieves efficient and high-fidelity medical image synthesis by utilizing diffusion models and adversarial learning. Li et al.\cite{li2023zero} propose a zero-sample medical image synthesis method based on frequency-guided diffusion models. It filters low and high-frequency information in the frequency domain to guide the generation of mid-frequency signals. Wolleb et al.\cite{wolleb2022diffusion} develop a semantic segmentation method based on diffusion model. It uses diffusion models to generate segmentation masks and improves the segmentation performance of medical images by modifying training and sampling schemes. Lyu et al.\cite{lyu2022conversion} propose an adaptation of denoising diffusion probability models and fractional matching models, employing four different sampling strategies for CT and MRI image synthesis. The results show that synthetic CT images outperform those created by CNN and GAN. Mao et al.\cite{mao2023disc} introduce a new multi-contrast MRI super-resolution method, DisC-Diff, which achieves super-resolution reconstruction of multi-contrast MRI using a conditional separation diffusion model. Yang et al.\cite{yang2023diffmic} propose DiffMIC, a medical image classification method based on diffusion networks. It improves medical image classification performance by a dual-granularity condition-guided and condition-specific maximum mean discrepancy (MMD) regularization. Zhao et al.\cite{zhao2023diffuld} introduce DiffULD, which uses a diffusion probability model for universal lesion detection. Experiments prove that DiffULD excels in locating lesions of different sizes and shapes, achieving performance comparable to state-of-the-art methods on both standard and enhanced DeepLesion datasets. Recognizing the potential of diffusion models in medical imaging tasks, we have applied diffusion models to dose prediction and conducted experiments on an internal dataset, the results of which surpass state-of-the-art methods.
\begin{figure*}
\centering
\includegraphics[width=1\linewidth]{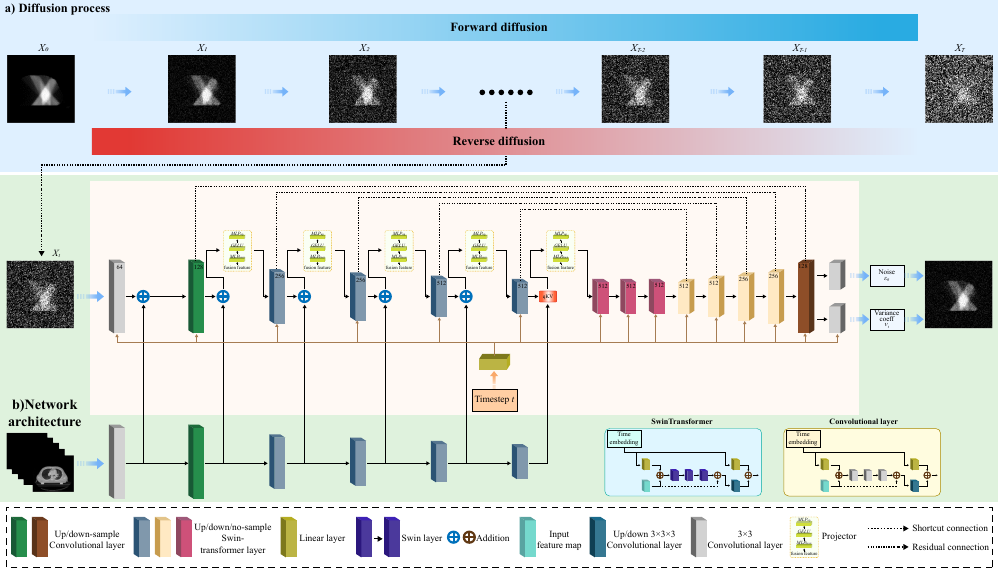}
   \caption{The method framework we propose is outlined, including (a) the forward noise injection and reverse denoising processes of the diffusion model and (b) the architecture diagram of the denoising network and structural encoding network.}
\label{fig2}
\end{figure*}

\section{Methodology}
Fig.~\ref{fig2} shows the overall network framework of the SP-DiffDose, where a) represents the forward noise addition and reverse denoising process of the diffusion model, and b) represents the proposed denoising network and structural encoder architecture. We add Gaussian noise to the dose distribution map t times during the forward diffusion process. In the reverse denoising process, we use a structural encoder to extract anatomical information, fuse the anatomical information with the noisy image, and input it into the denoising network to predict the noise, ultimately generating an accurate dose distribution map.
\subsection{Conditional DDPM}
In Conditional Denoising Diffusion Probabilistic Models (DDPM), the forward and reverse diffusion processes are two key steps. They respectively handle introducing noise into the data and reconstructing data from the noise space.

The forward diffusion process entirely transforms the data into noise incrementally through the Markov chain procedure. The definition of this process typically takes the form of a linear Gaussian transformation. For a given data $x_0$, the data at time step ($t$), $x_t$ can be generated through:
\begin{equation}
x_t = \sqrt{\bar{\alpha}_t} x_0 + \sqrt{1 - \bar{\alpha}_t} \epsilon,
\end{equation}
where $\epsilon$ is noise sampled from the standard normal distribution $\mathcal{N}(0, I)$. $\alpha_t$ is the variance scheduling parameter, typically $\alpha_t \in (0, 1)$. $\bar{\alpha}_t$ is the cumulative product of $\alpha_t$, defined as $\bar{\alpha}_t= \prod_{s=1}^t \alpha_{s}$.

This process starts from $t=0$ until $t=T$, at which point the data $x_T$ can be considered pure noise.

In the reverse diffusion process, we gradually transform the noisy data $x_T$ into the original data $x_0$. At each step, a neural network predicts the conditional probability distribution of the original data $x_{t-1}$, given the current noisy data $x_t$ and the conditioning information $y$. This prediction typically relies on estimating the following two values:

$\mu_{\theta}(x_t, t, y)$ - the conditional mean, representing the best estimate of $x_{t-1}$ given $x_t$ and the conditioning information $y$.

$\Sigma_{\theta}(x_t, t, y)$ - the conditional covariance, indicating the uncertainty of the prediction.

In practice, we do not directly predict $x_{t-1}$ but instead predict the noise $\epsilon$ and then use this noise along with the noise scheduling from the forward process to compute $x_{t-1}$. This strategy often enhances model stability and the quality of samples.

The following formula expresses the relationship between the predicted noise and the noise scheduling from the forward process:
\begin{equation}
x_{t-1} = \frac{1}{\sqrt{\alpha_t}} \left( x_t - \frac{1 - \alpha_t}{\sqrt{1 - \bar{\alpha}_t}} \epsilon_{\theta}(x_t, t, \phi(y)) \right),
\end{equation}
in this process, $\epsilon_{\theta}(x_t, t)$ is the noise predicted by the neural network, while $\alpha_t$ and $\bar{\alpha}_t$ are the variance scheduling parameters defined in the forward process.

Through this step-by-step reconstruction, we start from t=T and reverse to t=0, recovering a less noisy version of the data at each step until finally restoring the clean data $x_0$.
\subsection{Network Architecture}
\subsubsection{Structure Encoder}
It is challenging to preserve structural information by simply using the diffusion model for dose prediction, as the forward process only adds noise to the dose distribution map without incorporating structural priors, resulting in a significant discrepancy between the generated dose distribution map and the actual. Therefore, we design a structure encoder network to extract anatomical information from CT, PTV, and OARs to guide the denoising network in generating dose distribution maps. We input the structural image \(y \in \mathbb{R}^{(2+O) \times H \times W}\)(O represents the endangered organ, while H and W respectively stand for length and width) into the network for five times downsampling, obtaining the structure feature map \(f_i \in \mathbb{R}^{C_i \times H_i \times W_i}(i \in [1,5]\)) at each downsampling stage, and then fuse the structure feature map with the noise feature vector at the corresponding downsampling layer. First, we input \(y\) into a convolutional layer with a kernel size of 3$\times$3 and stride of 1 to learn early features. Then, it is input into a Conv-ResNet block containing three convolutional blocks and one downsampling convolutional block to learn early local features from the relatively high-resolution input. Then, four sequential downsampling Swin-ResNet blocks are applied to extract the structural information. The Swin-ResNet block contains three SwinTransformer blocks and one downsampling convolutional block, learning global features from low-resolution features; all downsampling operations use residual connections to avoid gradient vanishing. The SwinTransformer block operates as follows: We first divide the input feature map into non-overlapping local windows of size W$\times$W. For each window, we calculate the self-attention scores of all pixel pairs. For each pixel $i$ within the window, we first calculate its query vector $Q_i$, key vector $K_i$, and value vector $V_i$. Applying linear transformations to the input features of pixel $i$ yields these vectors:
\begin{equation}
Q_i = W_q \cdot x_i,
\end{equation}
\begin{equation}
K_i = W_k \cdot x_i,
\end{equation}
\begin{equation}
V_i = W_v \cdot x_i,
\end{equation}
where $W_q$, $W_k$, and $W_v$ are the weight matrices of the query, key, and value, respectively, and $x_i$ is the input feature of pixel $i$.

Next, we calculate the attention score of pixel $i$ and all other pixels $j$ within the window. The following formula can represent this:
\begin{equation}
a_{ij} = \text{softmax}(Q_i \cdot K_j^T / \sqrt{d_k}) \cdot V_j),
\end{equation}
where $d_k$ is the dimension of the key and value vectors, this formula calculates the similarity between pixel $i$ and pixel $j$, and uses this similarity to weight the value vector of pixel $j$.

After calculating the attention scores of all pixel pairs, we apply these attention scores to all pixels within the window to calculate the output features of each pixel within the window. The following formula can represent this:
\begin{equation}
h_i = \Sigma_j a_{ij} \cdot V_j,
\end{equation}
where $\Sigma_j$ represents the sum over all pixels $j$ within the window. 
We cyclically shift the windows between consecutive SwinTransformer blocks to deal with the information isolation problem between different windows. Each pixel can interact with its larger surrounding context in multiple blocks.

Finally, we process the output of the SwinTransformer block through a multilayer perceptron (MLP), which applies a series of linear transformations and non-linear activation functions. It further enhances the model's representational power. The operation of the MLP can be precisely represented by the following formula:
\begin{equation}
y_i = W_2 \cdot \text{relu}(W_1 \cdot h_i + b_1) + b_2
\end{equation}
where $W_1$, $W_2$, $b_1$, and $b_2$ are the weights and biases of the MLP, $\text{relu}$ is the non-linear activation function, $h_i$ is the output feature of pixel $i$, and $y_i$ is the final output feature of pixel $i$.
\subsubsection{Denoising Network}
We design a denoising network to convert the noise added during the forward process back into the dose distribution map. The denoising network is split into an encoder and a decoder for predicting noise at every step. The encoder has an architecture similar to the structural encoding network. At each encoder layer, it fuses the anatomical information extracted from the structural encoding network. It allows the denoising network to predict the dosage value and accurately locate the dosage area. Except for the last two layers, which use cross-attention mechanisms for fusion, all other layers use element-wise addition. For the cross-attention mechanism, we perform linear transformations on the structural feature mapping $A$ and the noise feature mapping $B$ to generate Query$(Q)$, Key$(K)$, and Value($V$) matrices. Specifically, the transformations are defined as follows:
\begin{equation}
Q_A = A W_Q^A,
\end{equation}
\begin{equation}
K_B = B W_K^B,
\end{equation}
\begin{equation}
V_B = B W_V^B,
\end{equation}
where $W_Q^A$, $W_K^B$, and $W_V^B$ are learnable weight matrices used to transform the original feature mappings $A$ and $B$ into their respective $Q$, $K$, and $V$ representations. Next, we calculate the attention scores between the queries $Q_A$ of $A$ and the keys $K_B$ of $B$. The scaled dot-product attention mechanism achieves it, as defined by the equation:
\begin{equation}
AttentionWeights = \text{Softmax}\left(\frac{Q_A K_B^T}{\sqrt{d_k}}\right),
\end{equation}
here, $d_k$ is the dimension of the key vectors used to scale the dot product to prevent the vanishing gradient problem. The Softmax function ensures that all attention weights sum up to 1, allowing the model to focus on the most relevant parts of feature mapping $B$. Applying the obtained attention weights to the values  $V_B$ produces a weighted feature representation. The output of this step is the final result of the cross-attention mechanism, calculated as:
\begin{equation}
Output = AttentionWeights \times V_B,
\end{equation}
here, the output represents the feature mapping with fused information from $A$ and $B$.

Then, two intermediate SwinTransformer blocks (without downsampling or upsampling) are connected to calculate global features further. The decoder replaces the Downblock in the encoder with Upblock, which is used to restore the feature representation extracted by the encoder to the original image size, finally getting the dosec distribution map. The time step uses sinusoidal embedding (SE)\cite{ho2020denoising} to encode, then the encoded time step embedding is input to all blocks for further calculation.

\subsubsection{Projector}
To efficiently combine features, we have developed a projector, the structure of which is depicted in Fig~\ref{fig2}. Our aim is to maintain the simplicity and efficiency of the projector. As such, our projector is composed of merely two multi-layer perceptrons (MLPs) and a Gaussian error linear unit (GELU) activation function.

Specifically, in the encoder part, we add the projector after the fusion of each layer of noise feature vectors and structural feature vectors. Expressing the equation as follows:
\begin{equation}
F_o=MLP_{up}(GELU(MLP_{down}(F_i))),  
\end{equation}
here, $F_o$ denotes the output features of the projector, and $F_i$ denotes the input features to the projector. $ MLP_{down}$ is a down-projection layer among all projectors, projecting features to a lower dimension, while $MLP_{up}$ is an up-projection layer among all projectors used for matching the dimensions of transformer features.
The purpose of this projector layer is to transform the input features more flexibly and expressively, which can enhance the denoising network's capability to eliminate noise while preserving the significant information in the source data.
\begin{table*}[ht]
\centering
\caption{Quantitative comparison results with state-of-the-art methods in terms of Dose Score, DVH Score and HI. The best results are highlighted in bold.}
\begin{tabularx}{\textwidth}{XXXXXXX}
\toprule
 & {Dose Score$\downarrow$} & {DVH Score$\downarrow$} & {HI$\downarrow$} \\
\midrule
{Cascade UNet} & 4.630$\pm$2.161 & 3.618$\pm$0.778 & 0.594$\pm$0.150\\
{HD UNet} & 4.172$\pm$1.749 & 3.195$\pm$0.608 & 0.560$\pm$0.122\\
{DiffDP} & 2.120$\pm$1.225 & 1.858$\pm$0.292 & 0.334$\pm$0.101\\
{SP-DiffDose} &\textbf{1.901}$\pm$\textbf{1.087} & \textbf{1.533}$\pm$\textbf{0.241} & \textbf{0.278}$\pm$\textbf{0.073}\\
\bottomrule
\end{tabularx}
\label{table1}
\end{table*}
\section{Experiments and results}
\subsection{Dataset and evaluation metrics}
The performance of SP-DiffDose is evaluated using a proprietary chest tumor dataset from Sichuan University West China Hospital, consisting of 300 treated patients. The dataset for each patient incorporates CT images, PTV segmentation, OARs segmentation, and the dose distribution from clinical plans, with OARs including the heart, lungs, and spinal cord. We randomly split the 320 patients into a training set of 220 patients, a validation set of 20 patients, and a test set of 80 patients. During training, we segment each 3D image into consecutive 2D slices, then resize all images to $256\times256$. Finally, we select only the images with dose information as input to the network. 

It is essential to use appropriate metrics to quantify the disparities between predicted and actual dose, the consistency of Dose-Volume Histogram(DVH) curves, and the uniformity of dose distribution. This study employs the following three primary metrics to evaluate the model's performance:

Dose Score quantifies the average relative deviation of the model from actual dose values. The calculation formula is as follows:
\begin{equation}
\text{Dose Score} = \frac{1}{n} \sum_{i=1}^{n} \frac{\hat{y}_i - y_i}{y_i},
\end{equation}
here, $n$ represents the sample size, $\hat{y}_i$ denotes the model's predicted dose values for the $i$-th sample, and $y_i$ signifies the true dose values for the $i$-th sample.

DVH Score assesses the model's performance by computing the differences in DVH curves between predicted and actual doses. The specific calculation process involves determining dose statistics such as D1, D95, and D99 and then calculating the average differences among these statistics. The calculation formula is as follows:
\begin{equation}
DVH \,\,Score = \frac{1}{n} \sum_{i=1}^{n} |\hat{y}_i - y_i|,
\end{equation}
The Homogeneity Index (HI) is used to characterize disparities in dose uniformity between predicted and actual dose images. It computes pixel values' mean and standard deviation and quantifies uniformity by dividing the standard deviation by the mean. The calculation formula is as follows:
\begin{equation}
HI = \frac{\sigma}{\mu},
\end{equation}
in this formula, $\sigma$ represents the standard deviation of pixel values, and $\mu$ represents the mean pixel value.
\begin{figure*}
\centering
\includegraphics[width=1\linewidth]{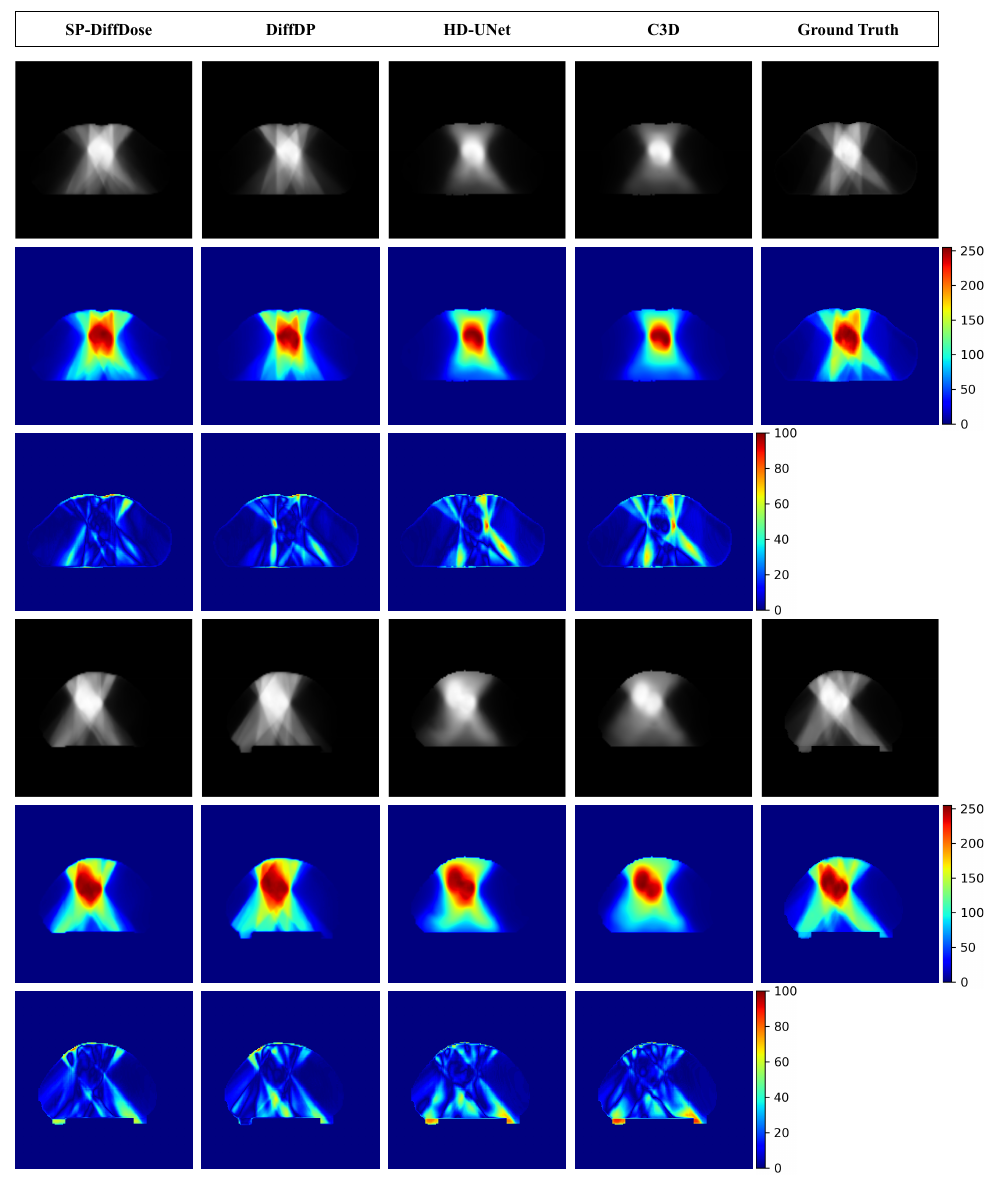}
   \caption{Two sets of visual comparisons are presented with state-of-the-art models. These include, from top to bottom, predicted dose distribution maps, heatmap depictions of dose distribution maps, and their respective difference maps. The final column signifies the ground truth.}
\label{fig3}
\end{figure*}
\subsection{Training Details}
We implement SP-DiffDose using PyTorch on the NVIDIA GeForce RTX 4090 GPU. The batch size is set to 8 throughout the experiment and utilizes the Adam optimizer. The entire model trains for 200 epochs. The learning rate is initialized at 1e-4 and linearly decays from 100 epochs until it reaches 0, aiming to expedite convergence and prevent getting stuck in local minima. Additionally, parameter T is set to 1000.

\subsection{Comparison with State-Of-The-Art Methods}
To validate the effectiveness of SP-DiffDose, we compare it with state-of-the-art (SOTA) methods in recent dose prediction research, including C3D, HD-UNet, and DiffDP. The quantitative comparison results are presented in Table~\ref{table1}, SP-DiffDose achieves a Dose Score of 1.901 Gy, a DVH Score of 1.533 Gy, and an HI index of 0.278. In comparison to C3D and HD-UNet, SP-DiffDose exhibits respective improvements of 58.94\% and 54.43\% in the Dose Score, 57.63\% and 52.02\% in the DVH Score, and an increase of 0.316 and 0.282 in the HI index. These results indicate that the diffusion model holds significant potential in the domain of dose prediction. Compared to DiffDP, SP-DiffDose shows increases of 10.33\% in the Dose Score, 17.49\% in the DVH Score, and 0.056 in the HI index, demonstrating that SP-DiffDose can more effectively predict dose.

Finally, we conduct paired t-tests to validate the significance of the results. The p-values between SP-DiffDose and other SOTA methods are all less than 0.01, indicating that the improvement in model performance is statistically significant.

In addition to the quantitative results, we present visual comparisons of our method with others in Fig.~\ref{fig3}. The dose distribution plots show that the dose prediction results obtained by C3D and HD-UNet closely resemble the ground truth within the PTV region. However, they overly smooth the prediction of high-frequency details, such as the beam direction and dose attenuation process, which is not conducive to subsequent improvements by physicists. DiffDP can predict details of the beam direction and dose attenuation process, but its accuracy in predicting doses to critical organs is relatively low. In contrast, SP-DiffDose not only accurately predicts the beam direction and dose attenuation process but also closely matches the ground truth for OARs dose predictions, achieving the best visual quality. Additionally, the dose difference maps between the third and fifth lines reveal that all methods yield predictions for the tumor dose distribution that closely match the ground truth. However, for regions near the tumor, the predictions by HD-UNet and C3D are less accurate, showing a more considerable discrepancy with the ground truth. Moreover, in some areas where the radiation beam initially enters, the predictions by DiffDP significantly differ from the ground truth. These areas are often OARs. Nevertheless, SP-DiffDose also shows a slight discrepancy with the ground truth, proving that SP-DiffDose can obtain rich high-frequency information and achieve the best visual results.

\begin{figure}
\centering
\includegraphics[width=1\linewidth]{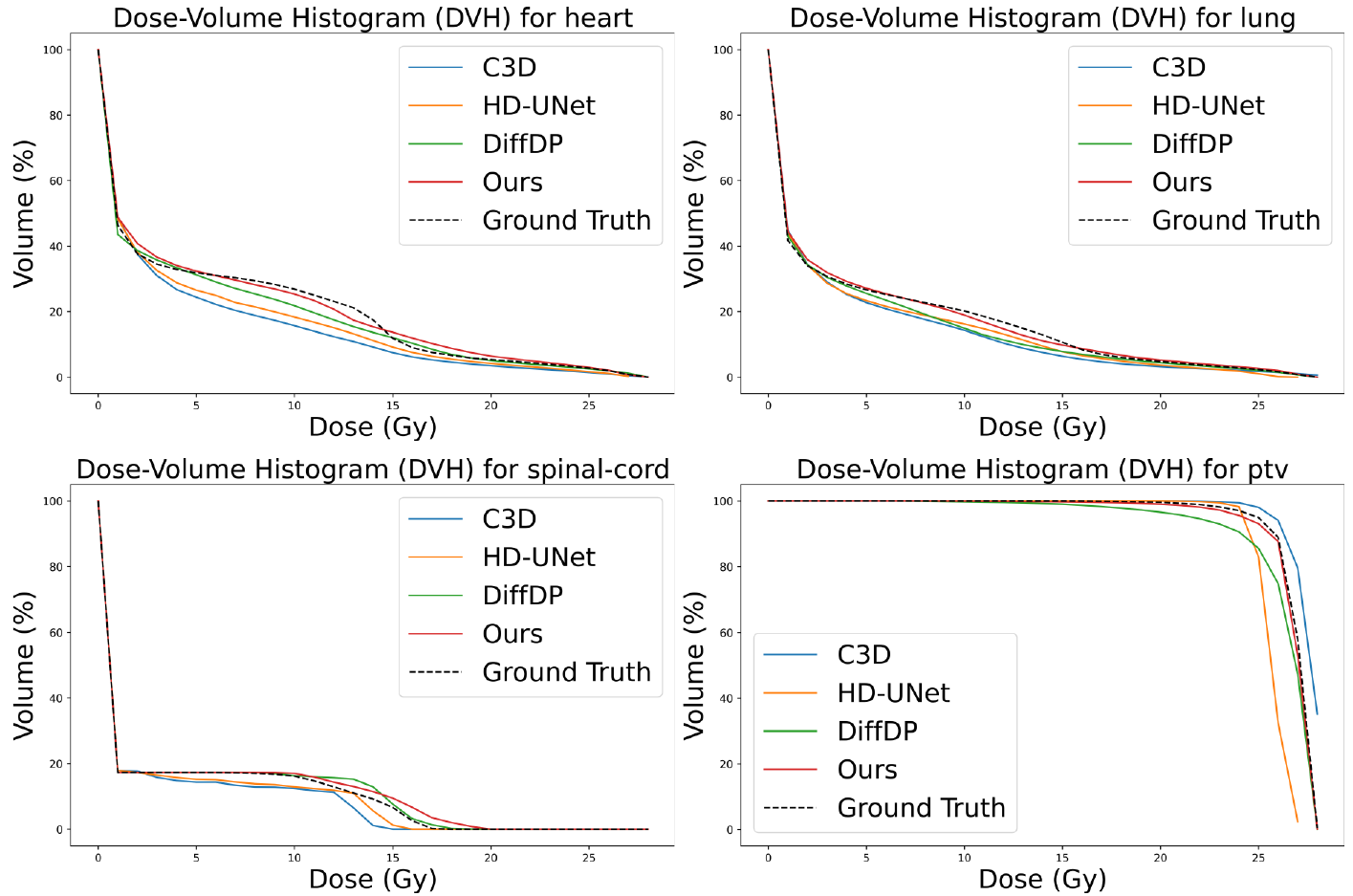}
   \caption{Visualize the DVH curves of our and SOTA methods, including the DVH curves for the PTV, Heart, Lung, and Spinal-cord.}
\label{fig4}
\end{figure}

\begin{table*}[ht]
\centering
\caption{Quantitative results of ablative studies on the chest tumor dataset. Mark the best results for each index in bold. "SwinTransformer" indicates whether to use the SwinTransformer in the network architecture. "Fusion" denotes the strategy of directly concatenating at the input layer or utilizing feature fusion at each network layer. "Projector" indicates whether to add a projector after feature fusion. The last row represents the proposed method.}
\begin{tabularx}{\textwidth}{XXXXXXX}
\toprule
SwinTransformer&Structure Encoder&Projector & {Dose Score}$\downarrow$ & {DVH Score}$\downarrow$ & {HI}$\downarrow$\\
\midrule
&&& 2.238$\pm$1.124 & 1.887$\pm$0.301 & 0.344$\pm$0.108\\
&$\checkmark$&& 2.219$\pm$1.189 & 1.878$\pm$0.315 & 0.342$\pm$0.108\\
&$\checkmark$&$\checkmark$& 2.120$\pm$1.225 & 1.858$\pm$0.292 & 0.334$\pm$0.101\\
$\checkmark$&&& 1.966$\pm$1.096 & 1.717$\pm$0.271 & 0.293$\pm$0.088\\
$\checkmark$&$\checkmark$&&1.922$\pm$1.078 & 1.610$\pm$0.248 & 0.282$\pm$0.078 \\
$\checkmark$&$\checkmark$&$\checkmark$&\textbf{1.901}$\pm$\textbf{1.087} & \textbf{1.533}$\pm$\textbf{0.241} & \textbf{0.278}$\pm$\textbf{0.073}\\
\bottomrule
\end{tabularx}
\label{table2}
\end{table*}
Furthermore, we calculate the volume percentages of different dose values on the PTV and OARs for the ground truth and the prediction results obtained by all compared methods. We have plotted the DVH curves, as shown in Fig~\ref{fig4}. SP-DiffDose's DVH curve closely matches the ground truth, providing evidence of the exceptional performance of SP-DiffDose.
\subsection{Ablation study}
Our approach consists of three core components: the SwinTransformer module, the structural encoder, and the projector. We conduct ablation experiments incrementally to demonstrate the contributions of the proposed modules. In Table~\ref{table2}, we present the results. Firstly, we validated the effectiveness of the SwinTransformer. Table~\ref{table2} shows whether or not utilizing the structural encoder and projector, the Dose Score, DVH Score, and HI all outperform the use of UNet as the feature extraction network. It demonstrates that the SwinTransformer can effectively capture features at different image scales. When integrating anatomical information extracted by the structural encoder with features extracted by the denoising network, the predictive performance surpasses that of directly concatenating noise and anatomical structure at the input layer, underscoring the efficacy of the structural encoder. 
\begin{table*}[ht]
\centering
\caption{Validate the impact of different fusion methods on the dose prediction results. 'concatenate' represents direct concatenation at the input layer, 'add-all' represents adding features at each layer for fusion, 'attn-all' represents using the cross-attention mechanism to fuse features at each layer, and 'attn-last2' represents using the cross-attention mechanism in the last two layers. In contrast, other layers use a direct addition fusion.}
\begin{tabularx}{\textwidth}{XXXXXXX}
\toprule
 & {Dose Score}$\downarrow$ & {DVH Score}$\downarrow$ & {HI}$\downarrow$\\
\midrule
Proposed/concatenate& 1.996$\pm$1.096 & 1.717$\pm$0.271 & 0.293$\pm$0.088\\
Proposed/add-all& 2.000$\pm$1.131 & 1.775$\pm$0.278 & 0.292$\pm$0.068\\
Proposed/attn-all& 6.108$\pm$1.920 & 3.181$\pm$0.639 & 0.527$\pm$0.159\\
Proposed/attn-last2&\textbf{1.901}$\pm$\textbf{1.087} & \textbf{1.533}$\pm$\textbf{0.241} & \textbf{0.278}$\pm$\textbf{0.073}\\
\bottomrule
\end{tabularx}
\label{table3}
\end{table*}

\begin{figure}
\centering
\includegraphics[width=1\linewidth]{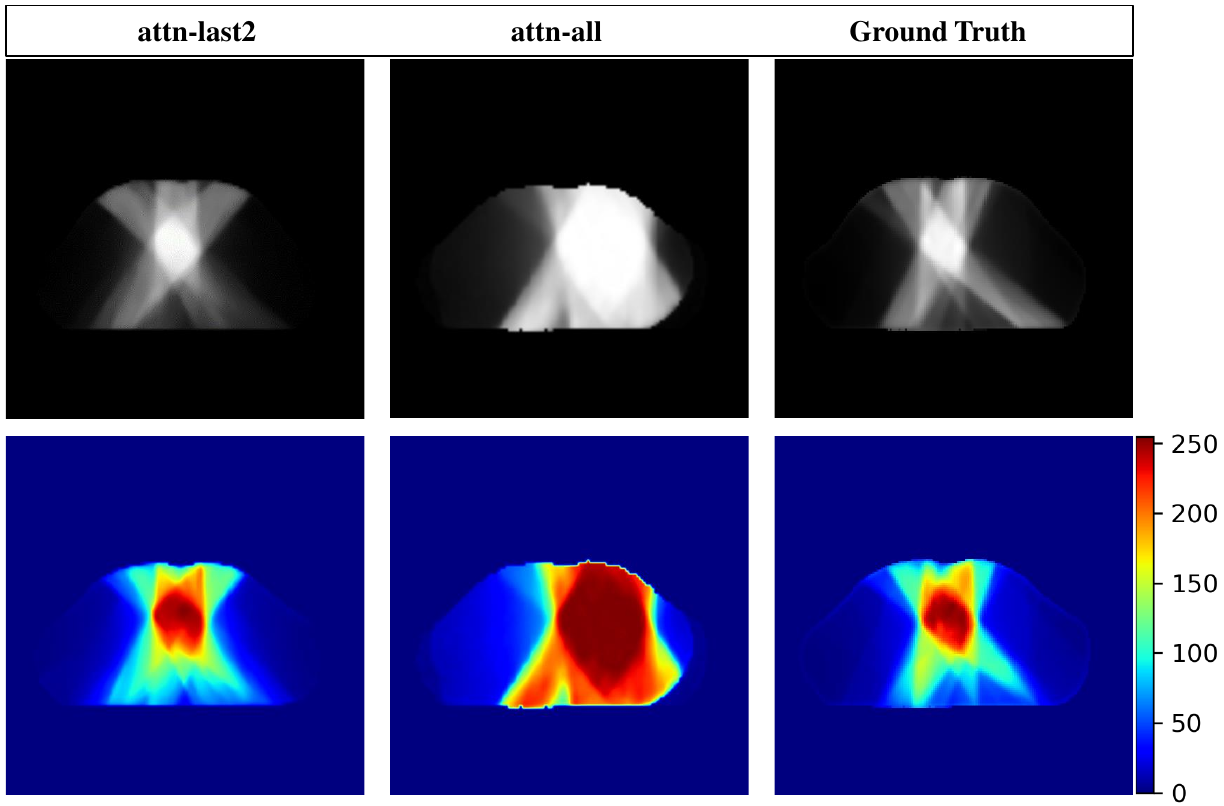}
   \caption{The visualization illustrates the distinction between employing cross-attention mechanisms for fusion at each layer of the network output and using cross-attention only in the last two layers for fusion. The last column represents the ground truth.}
\label{fig5}
\end{figure}

We further explored and validated the differences in predictive outcomes using various fusion methods, as shown in Table~\ref{table3}. When using the concatenate in network input, the dose prediction results are superior to those obtained with the add-all and attn-all fusion, particularly with the attn-all fusion. In the case of the attn-all fusion, the Dose Score, DVH Score, and HI index are 6.108 Gy, 3.181 Gy, and 0.527, respectively, lower than those of C3D and HD-UNet. We attribute this phenomenon to the fact that cross-attention can capture correlations between different image parts. However, in the early stages of the task, the features in different regions may need to be sufficiently consistent. It leads to the introduction of noise or inaccurate associations, thereby affecting the model's performance. As observed from visualization Fig.~\ref{fig5}, when using the attn-all fusion, the model predicts many high-dose regions, which deviate significantly from the ground truth. The attn-last2 fusion method also validates our analysis. When using attn-last2 for fusion, the Dose Score, DVH Score, and HI index are 1.901 Gy, 1.533 Gy, and 0.282, respectively, representing an improvement of 3.31\%, 10.71\%, and 0.015 compared to the concatenate fusion method. We further validate which layers benefit the model's dose prediction by using a cross-attention mechanism for feature fusion, as demonstrated in Table~\ref{table4}. The results indicate that employing a cross-attention mechanism in the last two layers yields optimal dose prediction results.
\begin{table*}[ht]
\centering
\caption{Validate the impact of using cross-attention mechanisms to fuse features at different layers on the dose prediction results. 'attn-lastx' (where x=[1,2,3,4]) represents using the cross-attention mechanism for fusion in the last x layers. In contrast, other layers use a direct addition fusion.}
\begin{tabularx}{\textwidth}{XXXXXXX}
\toprule
 & {Dose Score$\downarrow$} & {DVH Score$\downarrow$} & {HI$\downarrow$}\\
\midrule
attn-last4& 1.973$\pm$1.098 & 1.681$\pm$0.256 & 0.292$\pm$0.082\\
attn-last3& 1.965$\pm$1.086 & 1.663$\pm$0.263 & 0.291$\pm$0.092\\
attn-last1& 1.987$\pm$1.154 & 1.777$\pm$0.273 & 0.296$\pm$0.085\\
attn-last2&\textbf{1.901}$\pm$\textbf{1.087} & \textbf{1.533}$\pm$\textbf{0.241} & \textbf{0.278}$\pm$\textbf{0.073}\\
\bottomrule
\end{tabularx}
\label{table4}
\end{table*}

Finally, we assess the effectiveness of the projector. The experimental results in Table~\ref{table2}, specifically the fifth and sixth rows, illustrate the impact of the projector. When using the projector, the Dose Score, DVH Score, and HI indices improve from 1.922 Gy, 1.610 Gy, and 0.282 to 1.901 Gy, 1.533 Gy, and 0.278, respectively. This enhancement suggests that the projector is beneficial for dose prediction by optimizing the fused feature maps.

\section{Discussion}
In radiation therapy, accurate dose prediction is crucial for maximizing tumor control and protecting surrounding healthy tissues. However, formulating a high-quality radiation therapy plan remains a challenge due to the complex geometric shapes and positions of tumors. To address this issue, we propose a novel dose prediction model, SP-DiffDose, based on the diffusion model.

SP-DiffDose integrates a network architecture that combines convolution and SwinTransformer, capable of capturing both local and global features for more accurate predictions. Additionally, we develop a structural encoder that extracts available anatomical information from CT images, PTV, and OARs. We also introduce a projector to guide the denoising network for more precise predictions.

Compared with state-of-the-art methods such as C3D, HD-UNet, and DiffDP, SP-DiffDose outperforms three evaluation metrics: dose score, DVH score, and HI index. This performance indicates the significant potential of SP-DiffDose in the field of dose prediction. Particularly when compared with DiffDP, our model shows improvements in dose score, DVH score, and HI index.

In visual comparisons, SP-DiffDose not only accurately predicts the direction of the beam and the dose attenuation process but also closely matches the actual values in OARs dose prediction, achieving the best visual quality. Furthermore, the DVH curve of SP-DiffDose closely aligns with the actual values, demonstrating its excellent ability to calculate the volume percentage of different dose values on PTV and OARs.

SP-DiffDose demonstrates promising potential in dose prediction, outperforming existing state-of-the-art methods in quantitative and visual evaluations. Future work will focus on improving the model's performance and exploring its applicability in other fields. The results of this study provide new possibilities for the automation of radiation therapy planning, with the potential to improve treatment efficiency, reduce the workload of clinical professionals, and ensure treatment quality.

\section{Conclusion}
In this paper, we propose SP-DiffDose, a dose prediction method based on the diffusion model. SP-DiffDose gradually transforms the dose distribution map into gaussian noise during the forward process and generates the dose distribution map by gradually removing noise from the gaussian noise during the reverse process. To capture local and global features, we design a denoising network combining convolution and SwinTransformer. We use a structural encoder to extract patient anatomical information and design a projector to optimize the fused feature map to constrain the dose distribution of tumors and organs at risk. We conduct extensive experiments on an internal dataset of chest tumor to demonstrate the superiority of SP-DiffDose. Through our work, we use the dose prediction results as a good starting point for treatment planning, significantly reducing the time of the clinical workflow.

\section*{Acknowledgement}
This work was supported by 1.3.5 project for disciplines of excellence, West China Hospital, Sichuan University(20HXJS040).

\bibliographystyle{IEEEtran}
\bibliography{main}

\end{document}